\documentclass[12pt]{iopart}

\expandafter\let\csname equation*\endcsname\relax
\expandafter\let\csname endequation*\endcsname\relax

\usepackage{cite}
\usepackage[usenames,dvipsnames]{color}
\usepackage{bbm}
\usepackage{graphicx}
\usepackage{caption}
\usepackage{amsmath}
\usepackage{amssymb,amsthm}
\usepackage{hyperref}
\usepackage[normalem]{ulem}
\usepackage{xcolor}
\usepackage{mathtools}
\usepackage{colortbl}
\usepackage{comment}

\DeclarePairedDelimiter\floor{\lfloor}{\rfloor}

\newcommand{\eq}{\begin{eqnarray}} 
\newcommand{\en}{\end{eqnarray}}

\begin{document}

\title[Entanglement and fermionization of two distinguishable fermions]{Entanglement and fermionization of two distinguishable fermions in a strict and non strict one-dimensional space}
\author{Eloisa Cuestas$^{1,2}$, Mart\'in D. Jim\'enez$^{1,2}$, and Ana P. Majtey$^{1,2}$}

\address{$^1$ Universidad Nacional de C\'ordoba, Facultad de Matem\'atica, Astronom\'ia, F\'isica y Computaci\'on, Av. Medina Allende s/n, Ciudad Universitaria, X5000HUA C\'ordoba, Argentina}
\address{$^2$ Instituto de F\'{\i}sica Enrique Gaviola (IFEG), Consejo de Investigaciones Cient\'ificas y T\'ecnicas de la Rep\'ublica Argentina, C\'ordoba, Argentina}
\ead{eloisacuestas@unc.edu.ar}



\begin{abstract}
The fermionization regime and entanglement correlations of two distinguishable harmonically confined fermions interacting via a zero-range potential is addressed. We present two alternative representations of the ground state that we associate with two different types of one-dimensional spaces. These spaces, in turn, induce different correlations between particles and thus require a suitable definition of entanglement. We find that the entanglement of the ground state is strongly conditioned by those one-dimensional space features. We also find that in the strongly attractive regime the relative ground state is a highly localized state leading to maximum entanglement. Our analysis shows that in the strongly repulsive regime the ground state changes smoothly from a superposition of Slater-like states to a finite superposition of Slaters, this lack of accessible states yields to Pauli blocking as a strong signature of fermionization. Our results indicate that entangled states could be obtained in current experiments by reaching the non-interacting regime from the interacting regime. Entangled states could also be obtained when a state is brought from the interacting regime into the strongly repulsive regime by changing the scattering length near the confinement-induced resonance. Finally, we show that the first excited state obtained in the absence of interactions and the third excited fermionized state are maximally entangled. 
\end{abstract}





\section{Introduction}

Quantum entanglement is recognized by modern physics as one of the most intriguing and remarkable physical phenomena. Once established in the foundational discussions of quantum mechanics, its applications spread in many different directions, being perhaps the most compelling results those within the quantum information realm. Thus, in addition to being a fundamental concept in our understanding of Nature, quantum entanglement became a useful resource in many quantum tasks with the controlled manipulation of entangled states at the basis of several quantum information technologies \cite{nielsen_chuang_book,horodecki_2009}.
 
While the first progress regarding the theoretical development and the experimental implementation of entangled states in quantum information processing  arose in systems of photons, nowadays quantum information tasks can be also directly identified with ultracold atoms. Ensembles of entangled ultracold particles have been generated allowing for instance the implementation of highly sensitive interferometric measurements beyond the classical limit \cite{lucke_2011}. In the last twenty years ultracold atoms with magnetic-field tunable interactions have opened a completely new field connecting many different areas of physics. They constitute a key reference system in which some of the basic questions in many-body physics can be addressed and several methods can be tested \cite{zwerger_2012_book}. The interactions between trapped ultracold atoms cover many interesting collective quantum phenomena, ranging from Bose-Einstein condensation to recently observed fermionic superfluidity \cite{zwierlein_2005, zwierlein_2006, ku_2012, wang_2019}. Ultracold atoms also provide tunable systems in which few-body states can be prepared with high fidelities on demand \cite{serwane_2011}. The study of these few-particle systems allows for interesting connections between the physics of one or two bodies and the physics of many bodies \cite{liu_2009,liu_2010_PRA,liu_2010_PRB}, with strong consequences on our understanding of collective properties originated in the interactions between particles and in the statistics that rule them \cite{serwane_2011,zurn_2012,murmann_2015}.

Several interesting problems can be addressed with 1D systems which can be experimentally achieved by using ultracold confined atoms in strongly anisotropic potentials. This has made it possible to study Tonks-Girardeau and super—Tonks-Girardeau gases \cite{paredes_2004, haller_2010, girardeau_1960, astrakharchik_2005} and, from a more fundamental point of view, to go deeper into a key question of physics: how the dimensionality of a quantum system determines its physical properties \cite{sala_2013_PRL}. In particular, such one-dimensional confinement of few-fermion systems allowed to reach and study the fermionization of two distinguishable fermions \cite{zurn_2012,rontani_2012}. The existence of a fermionization regime was predicted by Girardeau as a mapping or one-to-one correspondence between a one-dimensional system of bosons with zero-range (and hard core) strongly repulsive interactions and an ideal Fermi gas. This mapping implies the fermionization of many properties of the bosonic system \cite{girardeau_1960}, the energy spectra are identical and the modules of the wave-function are the same. In other words, the bosons minimize their interaction energy by avoiding spatial overlap and exhibit fermionic properties leading to the formation of the Tonks-Girardeau (TG) gas. In addition, the phenomenon of fermionization has also been shown to be present in a system of two interacting distinguishable fermions \cite{girardeau_2010}.          

Fermionization of two distinguishable fermionic atoms confined in a strongly anisotropic trap is usually studied by considering an exactly solvable model (see Refs. \cite{busch_1998,avakian_1987}) consisting of two particles trapped in a one-dimensional harmonic potential with contact or zero-range interaction. The validity of this approximation was demonstrated in Ref. \cite{olshanii_1998}. The effective coupling strength between the fermionic atoms in two different hyperfine states (spin up and spin down) with s-wave interactions is controlled by tuning the scattering length via a Feshbach resonance \cite{chin_2010_review}, and the strongly repulsive regime related to fermionization is reached in the so called confinement-induced resonance (CIR) when the scattering length approaches the transversal confinement length \cite{olshanii_1998}. Even though the approximated model of two particles confined in a one-dimensional trap with delta type interaction has been widely studied and has quite known exact solutions, it continues providing new insights and allowing to describe and understand fascinating experiments like the ones of Refs. \cite{zurn_2012,rontani_2012}. Here, we focus on that model in order to study the entanglement, correlations, and fermionization in a system of two distinguishable confined fermions. The aim of this work is to analyze in detail the correlations present in the system for which we make use of two different ways of expressing its solutions. Such different representations of the system are extremely useful to describe its behavior across the full range of interaction, moreover, one of those representations explicitly exposes the fermionization phenomenon. As a plus, our discussion concerning how to deal with the notion of entanglement within these two approaches or representations shed light on some fundamental aspects of entanglement and the information reflected by each definition. Although the characterization and quantification of entanglement between distinguishable particles has received a lot of attention \cite{horodecki_2009},  the notion of entanglement in identical particle systems does not have a fully accepted definition \cite{schielmann_2001,pauskaukas_2001,ghirardi_2004,tichy_2011_JPB}. This is mainly due to the ambiguity that arises because the (anti)symmetry requirements  formally look like a non-local superposition. Our present analysis of an analytically solvable model indicates that some entanglement definitions proposed for indistinguishable particles provide relevant information regarding the correlations of the two-distinguishable particle system. In particular, we found that these entanglement definitions become  suitable tools to address and understand the different behaviors depicted by the system when considering a strict or rigorous one-dimensional space in which the particles can not pass trough each other or a more physically one-dimensional space as the ones obtained in current experiments which are actually performed in our three-dimensional world \cite{sala_2013_PRL}.

We show that the ground state of the system admits two representations that are related to two distinct types of one-dimensional spaces revealing different correlations between particles that require suitable definitions of entanglement. We discuss the proper definitions of entanglement depending on the information contained in each representation. A detailed analysis of the ground state shows that it is strongly conditioned by the one-dimensional space particularities. We show that in the strongly attractive regime the relative ground state is a highly localized state depicting maximum entanglement, while in the strong repulsive regime it changes smoothly from a superposition of Slater-like states to a finite superposition of Slater terms. This lack of accessible states for strongly repulsive interactions yields to Pauli blocking as a strong signature of fermionization. We also found that when reaching the non-interacting or strongly repulsive regime from the interacting regime the system selects entangled states over the non-entangled states expected by looking for solutions of the non-interacting or the infinitely repulsive system. 

The structure of the paper is as follows. In section \ref{sec_entanglement} we review the measures of entanglement for pure states, with emphasis on appropriate measures for systems composed of two identical fermionic particles and a discussion on the definitions made in order to consider only correlations beyond the wave-function antisymmetry. The model is reviewed in Sec. \ref{sec_model}. Section \ref{sec_gs} is devoted to the ground state behavior, two representations are introduced and a detailed study of the entanglement obtained with both representations is given. We discuss the strongly attractive regime and show that in this case the ground state presents maximum entanglement in Sec. \ref{sec_BEC}. In Sec. \ref{sec_unitarity} we show that entangled states could be obtained by experimentally reaching the non-interacting regime from the interacting regime. The strongly repulsive regime is discussed in Sec. \ref{sec_fermionization}, we show that the fermionized states can be written as a combination of Slater terms and calculate their entanglement showing that in current experiments maximally entangled states could be obtained in the CIR. Finally, a summary and conclusions are given in Sec. \ref{sec_concl}.


\section{Two-particle entanglement}
\label{sec_entanglement}

The experimental realization of the addressed system involves Feshbach tunable interactions between two particles of different species \cite{zurn_2012}, i.e. distinguishable particles. By using the center of mass and relative coordinates, the wave-function factorizes and the analytical solutions for the fundamental and excited states can be found for the whole range of interaction. In other words, the center of mass and relative variables define a bipartition of the Hilbert space in which the state is separable that is, the quasiparticles with center of mass and relative coordinates are not entangled. On the contrary, in the particle position space the interaction among particles induces quantum correlations between them meaning that when the bipartition is $H=H_1\otimes H_2$ with $H_i$ the Hilbert space of the $i-$th physical particle, the states are generally entangled. The resulting states depend on the relative distance $|r_1-r_2|$, where $r_{1(2)}$ is the spatial coordinate associated with particle $1(2)$. In Sec. \ref{sec_gs} we show that the ground state admits a representation as a sum of antisymmetric Slater-like determinants in terms of a new set of spatial coordinates defined as $r_>$ and $r_<$. The emergence of this alternative decomposition of the wave-function in terms of antisymmetric Slater (which for fermions is usually associated with indistinguishablility) is related to the lack of information in this new representation: we have one particle on the left (right) but we do not know whether the left (right) side particle with coordinate $r_<$ ($r_>$) is the particle associated with the standard spatial coordinate $r_1$ or the one represented by $r_2$. The wave-function is then written in terms of the coordinates $r_{>}$, $r_<$ which are not associated with one of the particles but with the right or left position of each of them. If the system under consideration is strictly one-dimensional and the particles cannot pass through each other, the variable $r_{<(>)}$ will be associated with only one of the particles with coordinates $r_{1(2)}$, in which case the (anti)symmetric character of the wave-function cannot be associated with a lack of the system's state information. In order to explore appropriately the correlations of the systems in both representation we use the available approaches for entanglement of systems formed by distinguishable and indistinguishable constituents. 

Due to all the above considerations, we devote the present section to the basics of entanglement in both, distinguishable and indistinguishable-fermion systems. In particular, we introduce the corresponding fermionic entanglement measures by briefly reviewing their distinguishable-particle counterpart. Let us start by considering a quantum system constituted by two distinguishable subsystems $A$ and $B$. Let $H_A$ and $H_B$ be the Hilbert spaces that describe the subsystems $A$ and $B$ respectively, then, the Hilbert space associated with the composite system has the form of the tensor product $H_{AB} = H_A \otimes H_B$. While pure states of such a composite system are separable (factorizable) whenever they can be written as, $\psi = \psi_A \otimes \psi_B$, entangled states cannot be written as the product of pure states corresponding to each subsystem. It should be noted that when we talk about entanglement, it is crucial to define which bipartition of the Hilbert space is being considered, that is, to define the entangled (or non entangled) subsystems.  

\subsection{Marginals mixedness degree}

As mentioned above, in the case of factorizable states each subsystem is characterized by a pure state. In contrast, when we have an entangled state it is impossible to assign an individual pure state to each subsystem, the subsystems are in mixed states and must be described with the formalism of density matrices. This is the reason why the degree of mixedness of the marginal density matrices constitutes a quantitative measure of the amount of entanglement of the pure state of the $A+B$ system. The larger the mixing of the subsystems, the greater the entanglement of the pure (global) state. In consonance with this, maximally entangled states are those with maximally mixed marginals ($\rho_{A(B)} \propto \mathbb{I}$, where $\rho_{A(B)}$ denote the marginal density matrix of subsystem $A(B)$). There are several ways to quantify the mixedness degree of the marginal density matrices and, consequently, the amount of entanglement associated with the whole pure state of the composite system. A standard measure of entanglement is based on the von Neumann entropy of any of the marginal density matrices $\rho_{A(B)}$. Let $\rho$ be for instance  $\rho_A$, then, the von Neumann entropy reads 

\eq
\label{eq_SvN}
S[\rho]= - \text{Tr}\, [ \rho \log \rho ] . 
\en

\noindent The degree of mixedness can also be quantified by the linear entropy $S_L$ of either one of the two reduced matrices, given by


\eq
\label{eq_SL}
S_L[\rho] = 1- \textrm{Tr}\,\rho^2.
\en

In the case of systems whose constituent subsystems or particles are identical, it is physically problematic (at least from the conceptual and fundamental perspective) to deal with the marginal states that describe each one of the particles. Due to the (anti)symmetrization of the wave-function, the state space of a system of indistinguishable particles is strictly restricted to a subspace of the tensor product Hilbert space. Thus, the Hilbert-space structure of indistinguishable particles no longer represents a physical partition into subsystems. As a consequence, the entanglement definition and entanglement measures ought to be adapted and addressed in such a way that they take into account the correlations between parties related to this quantum resource. Both measures, the one defined in terms of the von Neumann entropy and the one based in the linear entropy admit a natural generalization to quantify entanglement in systems of two identical particles \cite{plastino_2009_epl, manzano_2010, lopez_rosa_2015, tichy_2011_JPB}. One of the main features of these measurements is that correlations due only to the (anti)symmetry of the identical particle states do not contribute to the amount of entanglement. Thus the entanglement  corresponds to the quantum correlations exhibited by the state above the minimum one that satisfies the (anti)symmetric requirements of the wave-function. The latter is in agreement with the fact that the separability of a Slater determinant is consistent with the possibility of assigning a complete set of properties to each one of the subsystems, even when we cannot say which one of them because of their identical nature \cite{ghirardi_2004,tichy_2011_JPB}. In this line, elementary Slater determinants describing fermions are the analogues of product states in systems made up of distinguishable particles. Adequate generalizations of Eqs. (\ref{eq_SvN}) and (\ref{eq_SL}) lead to the following fermionic entanglement measures,

\eq
\label{eq_SvNf}
S^{f}[\rho] =- \text{Tr} \,[\rho \log(\rho) ]- \log(2) ,
\en

\noindent and

\eq
\label{eq_SLf}
S^{f}_L[\rho] = 1- 2\, \textrm{Tr}\,\rho^2 \,
\en

\noindent where $\rho$ is the reduced density matrix of the fermionic system \cite{schielmann_2001,pauskaukas_2001}. Throughout the manuscript we use $S_L$ to denote both the linear entropy and its redefinition made in order to consider correlations beyond the antisymmetric character of the wave-function (see Eqs.~\eqref{eq_SL} and ~\eqref{eq_SLf}). To avoid confusion, the use of $S_L$ is followed by a comment or the corresponding definition.

\subsection{Canonical decomposition, ranks, and numbers}
\label{subsec_shcmidt}

The characterization of entanglement for systems with distinguishable constituents is also achieved by the Schmidt decomposition through the Schmidt rank. A fermionic analogue to the Schmidt rank, which classifies entanglement in bipartite systems of identical fermionic particles is the so-called Slater rank. This quantity is associated to the fermionic Schmidt (or Slater) decomposition of the fermionic state.  Let the exact two-particle wave-function be $\Psi(x_1,x_2)$. The Schmidt/ Slater (or canonical) decomposition of $\Psi$ is characterized by the sum over a single-particle index

\eq
\label{eq_schmidt}
\Psi(x_1,x_2)=\sum_{j}\sqrt{\lambda_j}\,\Phi_j(x_1,x_2),
\en

\noindent where $\Phi_j(x_1,x_2)$ is a product of specific single-particles orbitals ($\varphi_j(x_1) \zeta_j(x_2)$, where $\lbrace \phi_j \rbrace$ and  $\lbrace \zeta_j \rbrace$ are orthonormal bases of $\mathcal{H}_1$ and $\mathcal{H}_2$ respectively) for distinguishable particles, and a Slater determinant of orthonormal single-particle orbitals for fermions ($[\varphi_{2j-1}(x_1)\varphi_{2j}(x_2)-\varphi_{2j}(x_1)\varphi_{2j-1}(x_2)]/\sqrt{2}$). This representation is unique and the Schmidt/ Slater rank is defined as the number of nonvanishing coefficients in the decomposition \cite{schielmann_2001,pauskaukas_2001}. A state is entangled if and only if its Schmidt/ Slater rank is strictly greater than one. The normalization of $\Psi$ leads to the condition $\sum_j \lambda_j = 1$, therefore, each coefficient $\lambda_j$ can be interpreted as a probability. Related to these decomposition and of particular interest in our future analysis, another entanglement measure can be defined as the inverse of the average probability $\sum_j \lambda_j^2$. The obtained measure, $\mathcal{K} = 1/\sum_j \lambda_j^2$ represents the effective number of involved single-particle states. The amount of entanglement can be roughly identified with the number of different functions $\Phi_{j}(x_1,x_2)$ needed in order to construct the exact two-particle wave-function \cite{grobe_1994}. In the case of the Schmidt decomposition $\lambda_j$ are the eigenvalues of the one-particle density matrix, while for the Slater or fermionic Schmidt decomposition each eigenvalue is two-fold degenerated and must be multiplied by 2 to obtain $\lambda_j$. Then, we can define the Schmidt number 

\eq
\label{eq_K}
\mathcal{K}= \frac{1}{\textrm{Tr}\rho^2} ,
\en

\noindent and the Slater number 

\eq
\label{eq_Kf}
\mathcal{K}^f = \frac{1}{ 2 \,\textrm{Tr}\rho^2} .
\en

\noindent Note that a single-Slater determinant corresponds to a single particle density matrix with only one two-fold degenerated eigenvalue equal to $1/2$, therefore, $\mathcal{K}^f=1$ represents a non-correlated state of identical fermions or an identical fermionic state with no correlations on the top of those required by the antisymmetry of the state. Also note that throughout the manuscript we use $\mathcal{K}$ to denote both the Schmidt and the Slater numbers  (see Eqs.~\eqref{eq_K} and ~\eqref{eq_Kf}). To avoid confusion, the use of $\mathcal{K}$ is followed by a comment or the corresponding definition.


\section{Two Particles in a Harmonic Trap, an Overview}
\label{sec_model}

Motivated by nuclear and high energy physics the spectroscopy of a singular oscillator was analytically addressed in Ref. \cite{avakian_1987}, where the author showed that a delta-shaped term radically  changes the oscillator spectroscopy. A decade after,  Busch et al. presented the exact solution to the problem of two ultracold particles confined in a harmonic trap interacting with a point-like potential of zero range within the shape-independent approximation \cite{busch_1998}. Although both works were devoted to theoretical derivations, the experimental progress in the control and manipulation of low-dimensional quantum systems with few particles made them extremely valuables tools for our understanding of the properties of ultracold atoms and Fermi gases.

Here, we consider two particles of mass $m$ in a one-dimensional harmonic trap of frequency $\omega$ with a zero-range potential interaction, 

\begin{small}
\eq 
\label{eq_H_full}
{\cal H} = -\frac{\hbar^2}{2m}\left(\frac{\partial^2}{\partial r_1^2}+\frac{\partial^2}{\partial r_2^2}\right) + 
\frac{m \omega^2}{2}\left(r_1^2 + r_2^2 \right)+ \gamma \delta(\vert r_1-r_2 \vert),
\en
\end{small}

\noindent where $r_1$ and $r_2$ denote the positions of the particles interacting via contact potential with strength $\gamma$. When introducing the center of mass $R = (r_1 + r_2)/2$ and relative coordinates $r = \vert r_1-r_2 \vert$ the above Hamiltonian decouples ${\cal H}={\cal H}_R+{\cal H}_r$, with ${\cal H}_R$ (${\cal H}_r$) describing the center of mass (relative) behavior, 

\begin{eqnarray}
\label{eq_HR_Hr}
{\cal H}_R & = & -\frac{\hbar^2}{2 m_R}\frac{d^2}{dR^2}+ \frac{m_R \omega}{2}R^2, \\
{\cal H}_r & = & -\frac{\hbar^2}{2 m_r}\frac{d^2}{dr^2}+ \frac{m_r \omega}{2}r^2 + \gamma \delta(r), 
\end{eqnarray}

\noindent where $m_R = 2m$ and $m_r = m/2$. This allows searching for solutions of the form $\psi =  \psi_R \, \psi_r$ with total energy $E = E_R + E_r$. The center of mass wave-function is then given by a one-dimensional oscillator state

\begin{small}
\eq 
\label{eq_psi_R}
\psi_{R}^n (r_1, r_2) = \left( \frac{2 m \omega }{\hbar \pi} \right)^{\frac{1}{4}} e^{- \frac{m \omega}{\hbar} \left( \frac{r_1 +r_2}{2} \right)^2} \frac{H_n\left( \sqrt{\frac{m \omega}{\hbar}} \frac{r_1 + r_2}{\sqrt{2}} \right)}{\sqrt{2^n n!}} ,
\en
\end{small}

\noindent with energy $E_R^n = \hbar \omega (n + 1/2)$. By expressing all the quantities in the characteristic units of the harmonic oscillator, the relative wave-function satisfies

\eq 
\label{eq_eq_relative}
0 =  \frac{d^2 \psi_r}{d x^2} + \left( \tilde{\lambda} + \frac{1}{2} - \frac{x^2}{4} \right) \psi_r - \tilde{\gamma} \delta(x) \psi_r ,
\en

\noindent where $x = \sqrt{m \omega / \hbar} \,r$, $\tilde{\lambda} = \epsilon_r - 1/2 = E_r/\hbar \omega - 1/2$, and $\tilde{\gamma} = \gamma \sqrt{m/ \omega \hbar^3}$. This last parameter is related to the two-body scattering length $a$ that can be obtained by solving the zero-energy scattering equation, which does not consider the confining potential \cite{lieb_2000, giorgini_2008_review}, 

\eq 
\label{eq_gamma_a}
\tilde{\gamma} =  - \frac{1}{\sqrt{\frac{m \omega}{\hbar}} a} = - \frac{1}{\tilde{a}}.
\en

\noindent In current low-dimensional experiments the interaction is controlled by means of Feshbach resonances. In the strongly repulsive regime a confinement-induced resonance arises when the three-dimensional scattering length approaches the characteristic length scale of the transversal confinement in a cigar-shaped potential \cite{haller_2010, zurn_2012}. This means that the parameter $\gamma$ (and therefore $\tilde{a}$) is related to the three-dimensional scattering length $a_{3D}$, 

\eq 
\label{eq_gamma_a_a3D}
\gamma = \frac{2 \hbar^2 a_{3D}}{m l_\perp^2} \frac{1}{1 - \vert \zeta(\frac{1}{2}) \vert \frac{a_{3D}}{l_\perp}} , 
\en

\noindent where $\zeta$ denotes the Riemann zeta function, and  $l_\perp=\sqrt{2\hbar/(m\omega_\perp)}$ is the characteristic length of the perpendicular confinement \footnote{In the experiments of Ref. \cite{zurn_2012} the cigar-shaped potential has a ratio between parallel and perpendicular confinement of about ten, i.e. $\omega_{\parallel}/\omega_{\perp} \sim 10$.}. Notice that Eq.~\eqref{eq_gamma_a_a3D} indicates that the one-dimensional coupling strength diverges when the scattering length approaches the characteristic length of the radial-confinement oscillator. 

The solutions of Eq.~\eqref{eq_eq_relative} are the well-known parabolic cylinder functions $D_{\tilde{\lambda}} (z)$ \cite{abramowitz_stegun_1964_book}. Following the procedure detailed in Ref. \cite{avakian_1987}, the obtained normalized relative wave-function reads 

\begin{small}
\begin{eqnarray}
\label{eq_psi_r}
\psi_{r}^{\tilde{\lambda}} (r_1, r_2) = & & \left( \frac{2 m \omega }{\hbar \pi} \right)^{\frac{1}{4}} \sqrt{\frac{\Gamma(-\tilde{\lambda})}{\Psi \left( \frac{1-\tilde{\lambda}}{2} \right) -\Psi \left( - \frac{\tilde{\lambda}}{2} \right) }}\,D_{\tilde{\lambda}} \left( \sqrt{\frac{m \omega}{\hbar}} \vert r_1 - r_2 \vert \right) ,
\end{eqnarray}
\end{small}

\noindent with $\Gamma$ and $\Psi$ denoting the Gamma and Digamma functions respectively. The relative energies depend on the interaction and satisfies

\eq 
\label{eq_e_r_gamma_a}
\tilde{\gamma} = - \sqrt{2} \frac{\Gamma \left( \frac{3}{4} - \frac{\epsilon_r}{2} \right) }{ \Gamma \left( \frac{1}{4} - \frac{\epsilon_r}{2} \right)} .
\en

\begin{figure}[tb]
\begin{center}
\includegraphics[height=0.35\columnwidth]{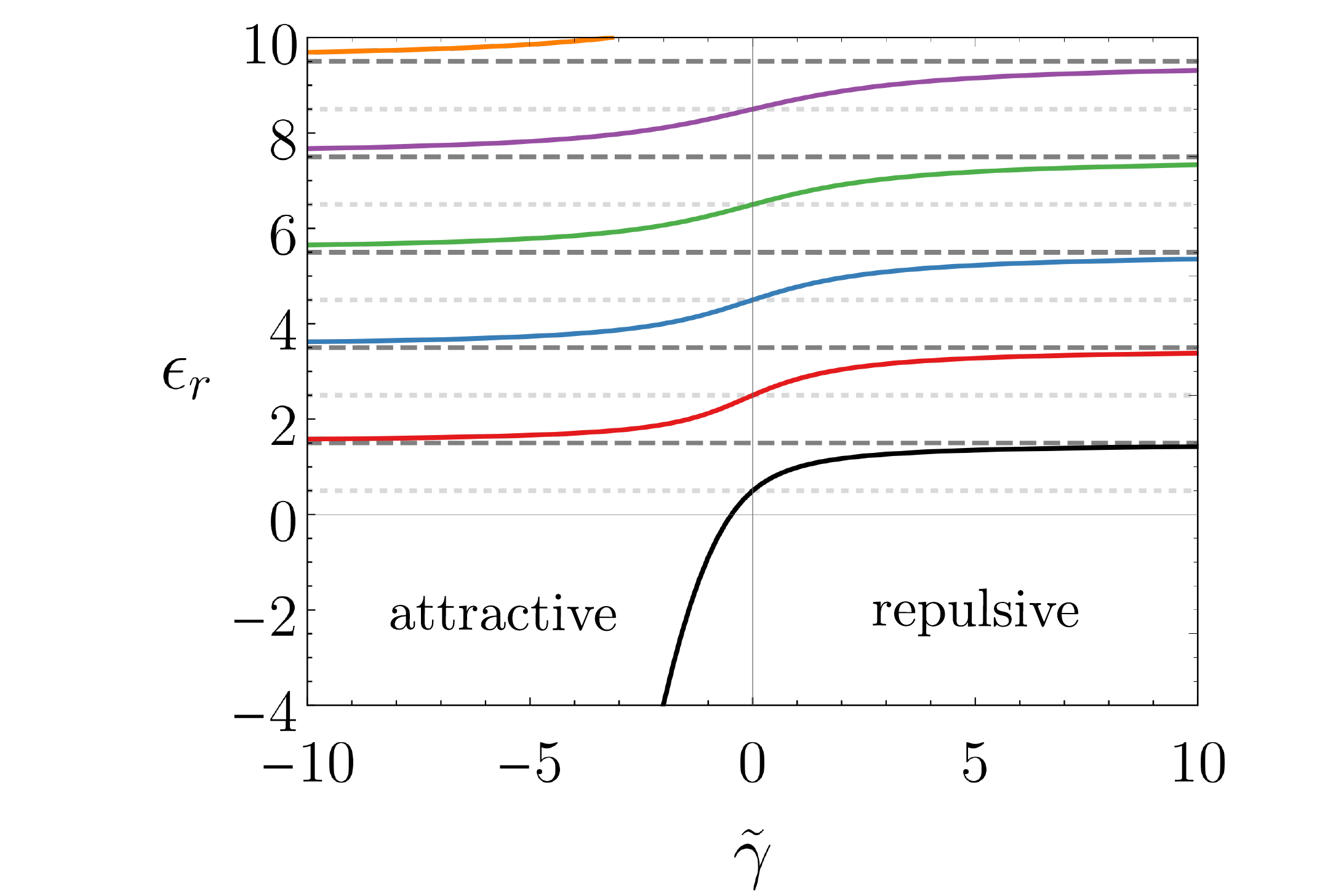}
\caption{Relative energy as a function of the interaction strength, related to the inverse of the scattering length by Eq.~\eqref{eq_gamma_a}. Light-gray dotted lines represent the zeros of the energy, while dark-gray dashed lines are the asymptotic values. The attractive and repulsive sides correspond to  $\tilde{\gamma}<0$ ($\tilde{a}>0$) and $\tilde{\gamma}>0$ ($\tilde{a}<0$).}
\label{fig_er}
\end{center}
\end{figure}

\noindent The relative energy spectrum is depicted in Fig. \ref{fig_er}. When the interaction is large and attractive the relative energy ground state is large and negative, when increasing $\tilde{\gamma}$ it increases reaching the asymptotic value of $\epsilon_r = 3/2$ for large repulsive interactions. The excited relative states present the same features and can be constructed by translating the first excited state by two energy units. In the repulsive (attractive) regime, these states have the asymptotic values $\epsilon_r = 3/2 + 2 n $ ($\epsilon_r = 3/2 + 2 (n-1) $) with $n=1,2,3,\ldots$, and vanish for $\epsilon_r = 1/2 + 2 n $, then the ground state can be associated with $n=0$ (with the caveat that this state presents only the repulsive asymptote). Notice that the non-interacting point in which the interaction changes from attractive to repulsive is given by $1/\tilde{a} = 0$ or equivalently by $\tilde{\gamma} = 0 $.  

In the deep repulsive regime the total wave-functions can be evaluated by taking $\tilde{\lambda} = \tilde{l} $ with $\tilde{l}=1$ for the ground state and $\tilde{l}=3,5,7,\ldots$ for the excited states, 

\begin{small}
\begin{eqnarray}
\label{eq_psi_BCS}
\psi_{sr}^{\tilde{\lambda} = \tilde{l}, \, n} (x_1, x_2) = & & \sqrt{\frac{m \omega }{\hbar \pi}} \frac{e^{-\frac{x_1^2+x_2^2}{2}}}{\sqrt{2^{\tilde{l}+n} \tilde{l}! n!}}\, H_n \left( \frac{x_1+x_2}{\sqrt{2}} \right) H_{\tilde{l}} \left( \frac{\vert x_1-x_2 \vert}{\sqrt{2}} \right) ,
\end{eqnarray}
\end{small}

\noindent where $x_i = \sqrt{m \omega / \hbar} \,r_i$ and $H_n(z)$ denotes the Hermite polynomial of order $n$. Deep into the attractive side $\tilde{\lambda} \ll -1$, the states can be approximated (see Ref. \cite{avakian_1987}) by 

\begin{eqnarray}
\label{eq_psi_BEC}
\psi_{sa}^{\tilde{\lambda}, \, n} (x_1, x_2) =  & & \sqrt{\frac{ m \omega }{\hbar}} \left( -\tilde{\lambda} \frac{2^3}{\pi} \right)^{\frac{1}{4}} e^{-\left( \frac{x_1+x_2}{2} \right)^2} \, \frac{H_n \left( \frac{x_1+x_2}{\sqrt{2}} \right)}{\sqrt{2^{n} n!}} e^{-\sqrt{-\tilde{\lambda}}\vert x_1-x_2 \vert} .
\end{eqnarray}

\noindent In fact, this last expression gives a very good approximation for $\tilde{\lambda} < -5$. 

Before moving to the next section we would like to mention that the key strategy used so far in order to obtain the exact wave-functions, i.e. the introduction of the center of mass and relative coordinates, relies heavily upon the harmonic confinement. Any deviation or deformation in the form of the trap leads directly to a coupling between these two degrees of freedom, resulting in transfer excitations between them. This can be very useful for the formation of bound pairs, as was experimentally observed in Ref. \cite{sala_2013_PRL}. 


\section{The Ground State: Representations and Entanglement}
\label{sec_gs}

This section is devoted to a detailed study of the ground state behavior. We show that the total ground state wave-function admits two different representations related to two different one-dimensional coordinates with physical meaning. These representations contain different physical information and reveal different features of the system. 

In what follows we present the main steps needed to obtain the two representations (details are given in \ref{sec_app_gs}). Expanding the parabolic cylinder function of Eq.~\eqref{eq_psi_r} as an Hermite series and using the identity 

\eq 
\label{eq_hermite_x+y}
H_n \left( \frac{z + w}{\sqrt{2}} \right) = \sum_{k=0}^{n} \frac{1}{2^{\frac{n}{2}}} \binom{n}{k} H_k(z) H_{n-k} (w)  ,
\nonumber
\en

\noindent as well as some symmetry properties of the Hermite polynomials \cite{gradshteyn_2007_book}, it is possible to see that the ground state of the system can be written as a bosonic-like expression, i.e a sum of Permanents (P) and product terms,

\begin{eqnarray}
\label{eq_gs_P}
\psi_{gs} (x_1, x_2) =  & & \sum_{n=1}^{\infty} \sum_{k=0}^{n-1} c^P(n,k,\tilde{\lambda}) P_{k, 2n-k} (x_1, x_2) 
\nonumber \\
& & +\sum_{n=0}^{\infty} c(n,\tilde{\lambda}) \phi_n(x_1) \phi_n(x_2) ,
\end{eqnarray}

\noindent or equivalently as a fermionic-like expression, i.e a sum of Slater-like terms (S),

\eq
\label{eq_gs_S}
\psi_{gs} (x_1, x_2) =  & & \sum_{n=0}^{\infty} \sum_{k=0}^{n} c^S(n,k,\tilde{\lambda}) S_{k, 2n+1-k} (x_1, x_2) ,
\en

\noindent where $\phi_n(z)$ denotes the $n$-th one-dimensional oscillator state. The coefficients of the bosonic-like expression are given by  

\begin{eqnarray}
\label{eq_c}
c(n,\tilde{\lambda}) = & &  \frac{\tilde{\lambda} 2^{\frac{\tilde{\lambda}+1}{2}}}{\Gamma(1-\frac{\tilde{\lambda}}{2})} \sqrt{\frac{\Gamma(-\tilde{\lambda})}{\Psi \left( \frac{1-\tilde{\lambda}}{2} \right) -\Psi \left( - \frac{\tilde{\lambda}}{2} \right) }} 
\, \frac{1}{\tilde{\lambda}-2n} \frac{(2n-1)!!}{2^n n!} ,
\end{eqnarray}

\noindent and

\eq
\label{eq_c_P}
c^P(n,k,\tilde{\lambda}) = \sqrt{2}\, c(n,\tilde{\lambda})\, \frac{n! (-1)^{k+n}}{\sqrt{(2n-k)! k!}},
\en

\noindent while the coefficients of the fermionic-like expression are

\begin{eqnarray}
\label{eq_z_S}
c^S(n,k,\tilde{\lambda}) =& & \frac{ 2^{\frac{\tilde{\lambda}}{2}+1}}{\Gamma(\frac{1-\tilde{\lambda}}{2})} \sqrt{\frac{\Gamma(-\tilde{\lambda})}{\Psi \left( \frac{1-\tilde{\lambda}}{2} \right) -\Psi \left( - \frac{\tilde{\lambda}}{2} \right) }} 
\nonumber \\
& &\times  \frac{1}{\tilde{\lambda}-(2n+1)} \frac{(2n+1)!!}{2^n} \frac{(-1)^{n+1+k}}{\sqrt{(2n+1-k)! k!}} .
\end{eqnarray}

\noindent The Permanents are indicated as

\eq
\label{eq_P}
P_{k, 2n-k} (x_1, x_2) = \frac{\phi_k(x_1) \phi_{2n-k}(x_2) + \phi_{2n-k}(x_1) \phi_k(x_2) }{\sqrt{2}},
\en

\noindent and the Slater-like terms are given by

\eq
\label{eq_S}
S_{k, 2n+1-k} (x_1, x_2) = \frac{\phi_k(x_<) \phi_{2n+1-k}(x_>) - \phi_{2n+1-k}(x_<) \phi_k(x_>) }{\sqrt{2}},
\en

\noindent with $x_>$ ($x_<$) being Max$(x_1,x_2)$ (Min$(x_1,x_2)$) in such a way that $\vert x_1-x_2 \vert = x_> - x_< $. 

The ground state is obtained by setting $n=0$ in Eq.~\eqref{eq_psi_R} and $-\infty < \tilde{\lambda} \leq 1$ in Eq.~\eqref{eq_psi_r} (first curve from bottom to top in Fig. \ref{fig_er}). Since it is a symmetric state under particle exchange, both representations have this symmetry. The representation given in Eq.~\eqref{eq_gs_P} has two clear symmetric contributions, one is a sum of Permanents while the other resembles a Schmidt decomposition (second right hand term). This bosonic-like representation includes symmetric states involving two single-particle states as well as double-occupancy states. On the other hand, the representation of Eq.~\eqref{eq_gs_S} is a Slater-like sum involving states in which the probability of finding both particles at the same position vanishes. Notice that the Slater-like contributions are not antisymmetric but symmetric (see Eq.~\eqref{eq_S}) under particle exchange, for this reason we refer to this states as Slater-like terms and not as proper Slaters.   

Due to the separability of the state in absence of interaction, for $\tilde{\gamma} = 0$ or equivalently $\tilde{\lambda} = 0 $ all the coefficients $c^P$ vanish as well as all the $c$ except the first one, then, in Eq.~\eqref{eq_gs_P} only remains the term $\phi_0(x_1) \phi_0(x_2)$. Moreover, when $\tilde{\lambda} \sim 0$ the second term of Eq.~\eqref{eq_gs_P} constitutes a very good approximation to the Schmidt decomposition of the state. For infinite repulsion ($\tilde{\lambda} = 1$) all the coefficients $c^S$ are null except the first one, therefore, the state is equivalent to a single Slater-like state $[\phi_0(x_<) \phi_{1}(x_>) - \phi_{1}(x_<) \phi_0(x_>)]/\sqrt{2}$. This means that the bosonic-like representation exposes that in the absence of interaction the system has a Schmidt rank equal to one, while the fermionic-like representation shows that for infinite repulsion the system has a Slater-like rank equal to one. This brings up a natural question: does this finite rank appear in a continuous or sudden way? In other words, do the coefficients vanish all at once or do they turn off slowly? To find an answer we need the Schmidt decomposition of the bosonic-like representation and the Slater decomposition of the fermionic-like representation. On the basis of Eqs.~\eqref{eq_gs_P} and \eqref{eq_gs_S}, and following the definitions and procedures described in Refs. \cite{schielmann_2001,pauskaukas_2001}, we calculate both decompositions by truncating the infinite $n$ sum of Eqs.~\eqref{eq_gs_P} and \eqref{eq_gs_S} up to a given $n=n_{max}$. 

\begin{figure}[tb]
\begin{center}
\includegraphics[height=0.35\columnwidth]{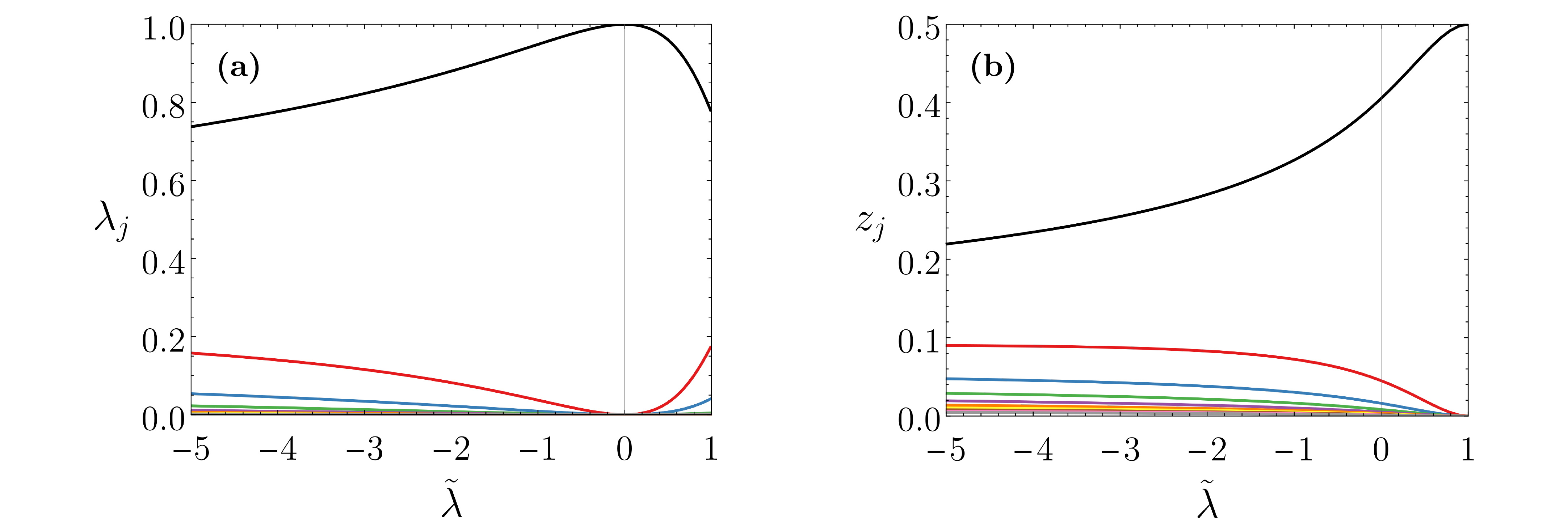}
\caption{Eigenvalues of the reduced density matrix as a function of the parameter $\tilde{\lambda}$. The first ten eigenvalues obtained with the bosonic-like representation (Eq.~\eqref{eq_gs_P}) are shown from top to bottom in panel (a), while the first ten two-fold degenerated eigenvalues obtained using the fermionic-like representation (Eq.~\eqref{eq_gs_S}) are depicted from top to bottom in panel (b). In the absence of interaction ($\tilde{\lambda}=0$) the Schmidt rank is equal to one (one coefficient), while the Slater rank is equal to one for infinite repulsion (for $\tilde{\lambda}=1$ there are two degenerated eigenvalues $z_j = 1/2$ corresponding to a single Slater-like term).}
\label{fig_schmidt_slater}
\end{center}
\end{figure}

Figure \ref{fig_schmidt_slater} depicts the eigenvalues of the reduced density matrix obtained by using the bosonic-like representation of Eq.~\eqref{eq_gs_P} and the fermionic-like representation of Eq.~\eqref{eq_gs_S}, which are denoted by $\lambda_j$ and $z_j$ respectively. Note that as stated in Sec. \ref{subsec_shcmidt} $\lambda_j$ is the $j$-th Schmidt coefficient while the two-fold degenerated $z_j$ must be multiplied by two in order to give the corresponding Slater coefficient. As expected, for $\tilde{\gamma} = 0$ ($\tilde{\lambda} = 0 $) the Schmidt rank of the bosonic-like representation is equal to one. When $\tilde{\lambda}$ goes to zero from the negative side, the larger Schmidt coefficient goes to one while all the remaining coefficients turn off slowly. For $\tilde{\lambda}>0$ the first Schmidt coefficient decreases while all the others turn on slowly. For infinite repulsion ($\tilde{\lambda} =1 $) the Slater rank is equal to one and there is a single (degenerated) non-vanishing Slater coefficient, which decreases for $\tilde{\lambda} <1$ while all the remaining coefficients slowly turn on. This behavior can be more directly appreciated by inspecting the Schmidt or Slater number ${\cal K}$ as a measure of the effective number of Schmidt or Slater modes \cite{law_2005_cob,law_2005_delta,pauskaukas_2001}, defined by $1/\text{Tr} \rho^2$ and $1/(2\, \text{Tr} \rho^2)$ respectively \cite{plastino_2009,pauskaukas_2001}. 

\begin{figure}[tb]
\begin{center}
\includegraphics[height=0.35\columnwidth]{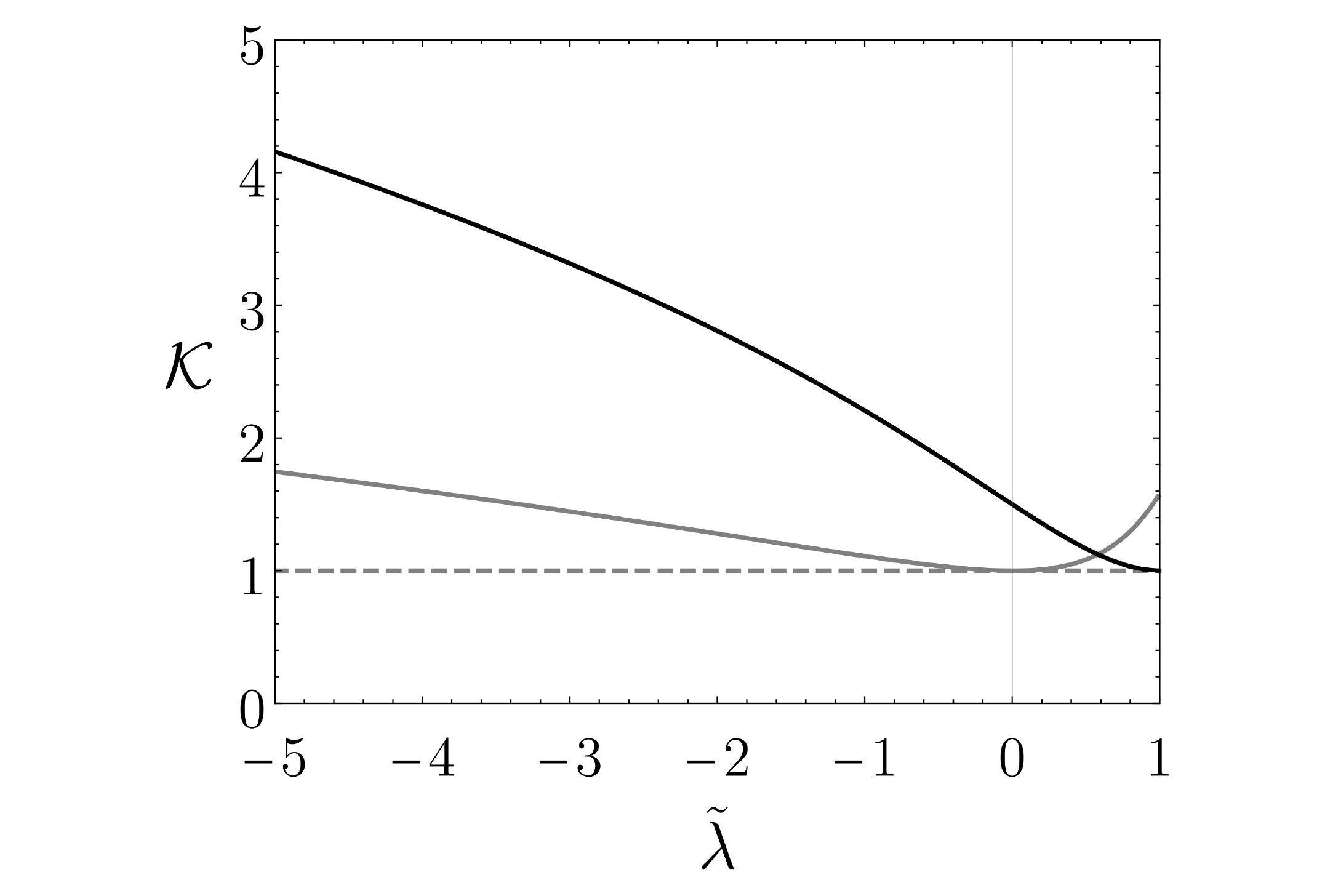}
\caption{Schmidt (gray solid line) and Slater number (black solid line) vs. $\tilde{\lambda}$. The values of $\tilde{\lambda}$ for which the Schmidt or Slater rank are equal to one are highlighted by the horizontal gray dashed line.}
\label{fig_K}
\end{center}
\end{figure}

Figure \ref{fig_K} shows the Schmidt (gray solid line) and Slater number (black solid line), exposing the values of $\tilde{\lambda}$ for which the Schmidt or Slater rank are equal to one. The Schmidt and Slater numbers have the same value for $\tilde{\lambda} = \tilde{\lambda}_{th} \sim 0.58$. When $ \tilde{\lambda}_{th} < \tilde{\lambda} < 1$ the number of effective modes contributing in the bosonic-like representation is larger than the number of modes involved in the fermionic-like representation. For $\tilde{\lambda} < \tilde{\lambda}_{th}$ the fermionic-like representation involves a larger number of modes than the bosonic-like representation. There are another two interesting issues here: first, for infinite repulsion the number of effective Schmidt modes is finite, and second, the number of effective Slater modes increases more quickly than the number of Schmidt modes. Summing up, the system has more accessible states in the fermionic-like representation than in the bosonic-like one. Regarding this point it is necessary to be very careful; for $\tilde{\lambda} < \tilde{\lambda}_{th}$ the system has more accessible two-particle modes in the fermionic-like representation (slater-like states) than in the bosonic-like one (product states $\phi_n(x_1) \phi_n(x_2)$). Since the Schmidt modes involve only one single particle state while the Slater modes involve two, and there are no repeated terms in the Slater decomposition, this means that the effective number of  accessible single particle states is quantified by $1/\text{Tr} \rho^2$ in both cases. From Fig. \ref{fig_K} it is easy to see that $1/\text{Tr} \rho^2 = 2\, {\cal K}^{\text{fermionic-like}}$ (twice the black solid line) is larger than $1/\text{Tr} \rho^2 = {\cal K}^{\text{bosonic-like}}$ (the gray solid line) for any value of $\tilde{\lambda}$. 

\begin{figure}[tb]
\begin{center}
\includegraphics[height=0.35\columnwidth]{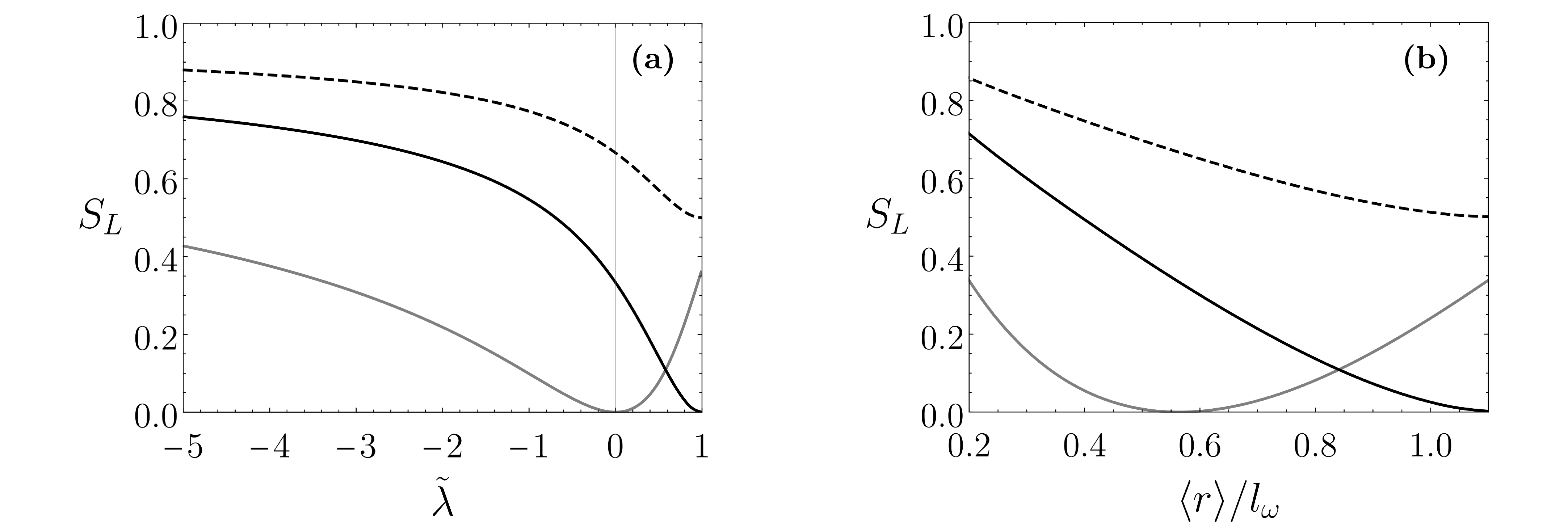}
\caption{(a) Entanglement as a function of $\tilde{\lambda}$ for the bosonic-like representation (gray line) and for the fermionic-like representation (black line). In the absence of interaction ($\tilde{\lambda} \sim 0$) the Schmidt rank goes to one and there are no quantum correlations, thus the bosonic-like entanglement vanishes. For a large enough repulsion ($\tilde{\lambda} \sim 1$) the fermionic-like entanglement vanishes exposing that there are no quantum correlations beyond the antisymmetry correlations (the Slater rank is equal to one). The entanglement obtained for the fermionic-like representation when considering correlations due to the antisymmetry of the wave-function is depicted as a black dashed line. This entanglement corresponds to a strictly one-dimensional space in which the particles can not pass through each other. (b) Entanglement as a function of the pair size divided by the characteristic length of the trap obtained for the same values of $\tilde{\lambda}$ as in the main plot.}
\label{fig_SL}
\end{center}
\end{figure}

Entanglement measures are computed straightforward from the Schmidt or Slater decomposition \cite{schielmann_2001,pauskaukas_2001}. In figure \ref{fig_SL} we show the linear entropy obtained for the bosonic-like representation (gray line) as well as the one obtained for the fermionic-like representation (black line), defined as $S_L = 1- \text{Tr} \rho^2$ and $S_L = 1- 2\, \text{Tr} \rho^2$ respectively. Once again the fermionic-like definition is made in such a way of considering correlations beyond antisymmetry \cite{plastino_2009}. As expected from our analysis of the Schmidt and Slater coefficients and numbers, the bosonic-like entanglement vanishes in the absence of interaction (the Schmidt rank is equal to one) while the fermionic-like entanglement vanishes for infinite repulsion (the Slater rank is equal to one). The entanglement increases when the interaction gets more an more attractive ($\tilde{\lambda} \ll -1$) and the pair size relative to the trap size decreases, see panel (b) of Fig. \ref{fig_SL}, where the entanglement as a function of the pair size divided by the characteristic length of the trap is depicted. This increase in the entanglement is also related with the increasing number of accessible one-particle states  quantified by $1/\text{Tr} \rho^2$, in consonance with the discussion presented in Ref. \cite{baccetti_2013}. This features are in complete agreement with the results obtained in Refs. \cite{chudzicki_2010, cuestas_2020}, in which the availability of enough space in the real space was related to the availability of sufficient space in the state space. 

Figure \ref{fig_SL} also shows the obtained entanglement for the fermionic-like representation when including the correlations due to the antisymmetric character of the wave-function (black dashed line), i.e. defined as $S_L = 1- \text{Tr} \rho^2$. This entanglement quantifies the correlations of two-particles in a strictly one-dimensional space in which the particles can not pass through each other and the same particle is always confined to be for instance in the same left or right side. In this sense the entanglement measure is strongly conditioned by the type of the one-dimensional space: in a strict one-dimensional space the particles have an intrinsic restriction which imposes an extra correlation leading to higher entanglement (black dashed line), while in a non strict one-dimensional space in which particles can occupy any position this extra spurious correlation must be discarded leading to a smaller entanglement (black line). Since the experimental realization of one-dimensional traps are actually three dimensional cigar-shaped potentials, the characteristics of those physical systems require the first definition of entanglement ($S_L = 1- 2\, \text{Tr} \rho^2$, black dashed line in Fig. \ref{fig_SL}), from now on we will refer to this as the fermionic-like representation entanglement.

We would like to make two final notes related to the presented results. First, the von Neumann and R\'enyi entropies show a very similar behavior to the linear entropy. In consonance with the results of Refs. \cite{franchini_2014,hamma_2013} for finite dimensional systems and in Refs. \cite{garagiola_2016} for continuous variables, the R\'enyi entropies present a non-analytical behavior in the values of $\tilde{\lambda}$ for which the Schmidt or Slater rank are equal to one. Second, although we focused on the study of the ground state, by using Eqs.~\eqref{eq_psi_R} and \eqref{eq_psi_r} as well as some properties of the Hermite polynomials it is possible to obtain general expressions equivalent to Eqs.~\eqref{eq_gs_P} and \eqref{eq_gs_S} for the excited states (see last paragraph of \ref{sec_app_gs}).  


\section{Deeply into the Attractive Regime: States and Correlations}
\label{sec_BEC}

For sufficiently strong attraction the two fermions present a molecular-like state and behave as a single entity. As mentioned in Sec. \ref{sec_model}, for $\tilde{\gamma} < - 2.13$ ($\tilde{\lambda} < -5$) the wave-function can be approximated by Eq.~\eqref{eq_psi_BEC}. Then, deep into the attractive side, i.e. for $\tilde{\gamma} \ll - 1$ ($\tilde{\lambda} \ll -1$), the squared modulus of the wave-function is

\eq
\label{eq_psi_BEC_full}
\left\vert \psi_{sa}^{\tilde{\lambda} \ll -1, n} \right\vert^2 = \left( 2^{\frac{1}{4}} \phi_n(\sqrt{2} R) \right)^2 \delta(r) ,
\en

\noindent where we used the identity $\lim_{\epsilon \to \infty} \sqrt{\epsilon} e^{-2 \sqrt{\epsilon} \vert z \vert }= \delta{(z)}$ \cite{avakian_1987}. The obtained states are a product of a highly localized state in the relative coordinate and a one-dimensional oscillator in the center of mass variable which gets more delocalized for higher $n$. 

The purity $P$ of the ground state ($n=0$) vanishes. This can be easily seen by considering that $P = \text{Tr} \rho^2 = \int dx \int dx^\prime (\rho(x,x^\prime))^2$ with $\rho(x,x^\prime)=\int dx_2 \, \psi(x,x_2) \psi(x^\prime,x_2)$, taking into account that the wave-function is real, and using a more suitable expression for the Dirac delta function. Therefore the entanglement measures based on the linear entropy studied in the previous section assume their maximum values (one) for $\tilde{\lambda} \ll -1$. 


\section{Absence of interaction and Correlations}
\label{sec_unitarity}

In the absence of interaction $\tilde{\gamma}=\tilde{\lambda}=0$, the total state is

\eq
\label{eq_psi_unitarity}
\psi^{\tilde{\lambda} = 0 , n}  = \sqrt{\frac{m \omega}{\hbar \pi}} e^{-\frac{x_1^2+x_2^2}{2}} \frac{H_n \left( \frac{x_1+x_2}{\sqrt{2}} \right)}{\sqrt{2^n n!}} ,
\en

\noindent which can be written as

\begin{eqnarray}
\label{eq_psi_unitarity_dec}
\psi^{\tilde{\lambda} = 0 , n}  = & & \sum_{k=0}^{\bar{n}} \sqrt{\frac{\binom{n}{k}}{2^{n-1}}} P_{k, n-k} (x_1, x_2) 
 + \, p(n) \sqrt{\frac{n!}{2^n}} \frac{1}{\left(\frac{n}{2}\right)!} \phi_{\frac{n}{2}}(x_1) \phi_{\frac{n}{2}}(x_2) ,
\end{eqnarray}

\noindent where $\bar{n} = n/2 -1$ and $p(n)=1$ for even $n$, while $\bar{n} = (n-1)/2$ and $p(n)=0$ for odd $n$. The permanent $P_{k, n-k} (x_1, x_2)$ fulfills

\eq
\label{eq_permanent_unitarity}
P_{k, n-k} (x_1, x_2) = \frac{\phi^+_{n,n-k} (x_1) \phi^+_{n,n-k} (x_2) - \phi^-_{n,n-k} (x_1) \phi^-_{n,n-k} (x_2) }{\sqrt{2}} ,
\en

\noindent with $\phi^+_{n,n-k} (y) = [\phi_n(y)+\phi_{n-k}(y)]/\sqrt{2}$ and $\phi^-_{n,n-k} (y) = [\phi_n(y)-\phi_{n-k}(y)]/\sqrt{2}$. The states $\phi_{n,n-k}^{+/-}$ are orthogonal to the $\phi_{n'}$ states of the second term because $k=0,1,\ldots,\bar{n}<n/2$, implying that $n' \neq n$ and $n' \neq n-k$. Since the indices appearing in Eq.~\eqref{eq_psi_unitarity_dec} are all distinct, the orthogonality between $\phi_{n,n-k}^{+/-}$ and $\phi_{n',n'-k'}^{+/-}$ is ensured. The orthogonality among $\phi_{n,n-k}^{+}$ and $\phi_{n',n'-k'}^{-}$ also holds due to the non-repetition of the indices except for $n=n'$ and $k=k'$, case in which the states are also orthogonal due to the different signs in the definitions of $\phi_{n,n-k}^{+/-}$. In the light of all this, for a given $n$ the states $\phi_{n,n-k}^{+/-}$ with $k=0,1,\ldots,\bar{n}$ and $\phi_{n/2}$ are the natural orbitals with associated occupations $\binom{n}{k}/2^n$ and $\binom{n}{n/2}/2^n$ respectively. The linear entropy of the single-particle reduced density matrix reads   

\eq
\label{eq_SL_unitarity}
S_L^{\tilde{\lambda} = 0 , n}  = 1- \left( 2 \sum_{k=0}^{\bar{n}} \frac{\binom{n}{k}^2}{2^{2 n}} + p(n) \frac{\binom{n}{\frac{n}{2}}^2}{2^{2 n}} \right) ,
\en
 
\noindent notice that the total energy of the state $\psi^{\tilde{\lambda} = 0 , n}$ is $E/\hbar\omega = \epsilon = n+1$ and therefore the entanglement can be written in terms of the total energy. 

\begin{figure}[tb]
\begin{center}
\includegraphics[height=0.35\columnwidth]{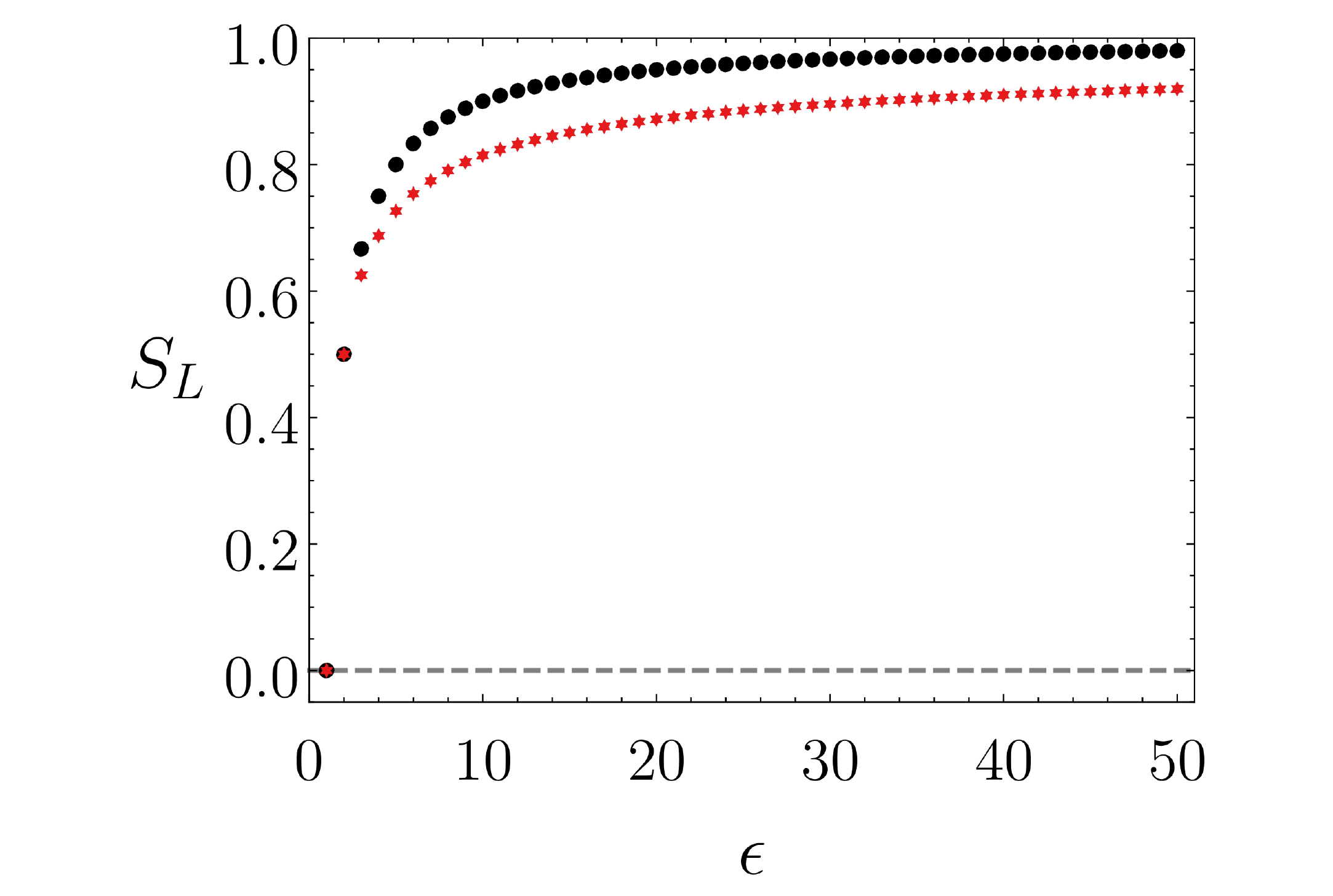}
\caption{Entanglement as a function of the total energy of the state in the absence of interaction (red stars). The amount of entanglement increases for increasing energy. The black dots give the upper bound of the entanglement, which is only reached by the first excited state. Notice that the entanglement vanishes for the ground state.}
\label{fig_SL_unitarity}
\end{center}
\end{figure}

Figure \ref{fig_SL_unitarity} shows the entanglement as a function of the total energy of the state (red stars). In the limit of vanishing interaction the ground state ($n=0$, $\epsilon =1$) is not entangled. The amount of entanglement is an increasing function of the total energy, since $\tilde{\lambda}=0$ these increasing correlations are provided by the center of mass state. For a given energy the entanglement is bounded by the entanglement of an equally weighted sum of the product states involved in the Schmidt decomposition of the total wave-function \cite{horodecki_2009,tichy_2011_JPB,nielsen_chuang_book}, represented in Fig. \ref{fig_SL_unitarity} by black dots,

\eq
\label{eq_SL_unitarity_bound}
\left(S_L^{\tilde{\lambda} = 0 , n}\right)^{bound} = 1- \frac{1}{\epsilon} ,
\en

\noindent notice that $ \epsilon = n+1 = g_{sp}$, with $g_{sp}$ being the number of available single-particle states. The bound given by Eq. (\ref{eq_SL_unitarity_bound}) is only reached by the first excited state ($\epsilon=2$ or $n=1$), while for $n \neq 1$ the entanglement remains under this bound. 

The results of the present section imply that entangled states could be obtained by experimentally reaching the non-interacting regime from the interacting regime with $\tilde{\lambda} \neq 0$. When solving Eq.~\eqref{eq_H_full} in the absence of interaction, the obtained states are $\phi_n(x_1) \phi_m (x_2)$ with energy $E = \hbar \omega (n+m+1)$. These states have zero entanglement. In contrast, bringing a state from the interacting regime $\tilde{\lambda} \neq 0$ to the non-interacting regime by changing the scattering length, leads to entangled states. Moreover, the first excited state obtained with this procedure would present the maximum entanglement for two distinguishable particles with two one-particle accessible states. 

The presence of excited spin up/down states exhibiting entanglement even in the limit of vanishing interaction has already been described in Refs. \cite{yanez_2010, tichy_2011_JPB}. Entangled states have also been obtained in this limit in Ref. \cite{majtey_2012}, where the entanglement between two electrons with a small arbitrary interaction was analyzed via a perturbative approach. The authors of Ref. \cite{majtey_2012} showed that the non-zero entanglement present in systems of interacting particles in the limit of vanishing interaction is due to the non-interacting Hamiltonian eigenstates selected by the interaction. Our results reinforce that entangled states can be obtained for vanishing interactions and agree with the notion that when reaching the non-interacting regime from the interacting one the system selects entangled states over non-entangled states.




\section{Deeply into the Repulsive Regime: Fermionization and Correlations}
\label{sec_fermionization}

The total wave-function in the strong repulsive limit is given by Eq.~\eqref{eq_psi_BCS}. In \ref{sec_app_slaters} we show that these states can be written as a combination of  Slater-like terms, 

\begin{small}
\eq
\label{eq_psi_BCS_n0}
\psi^{\tilde{\lambda} = \tilde{l}, \, n=0} (x_1, x_2) = & & \sqrt{\frac{\tilde{l}!}{2^{(\tilde{l}-1)}}} \, \sum_{l=0}^{\frac{\tilde{l}-1}{2}} \frac{(-1)^l}{\sqrt{(\tilde{l}-l)!l!}} S_{l,\tilde{l}-l} (x_1, x_2) ,
\en
\end{small}

\noindent and

\begin{small}
\eq
\label{eq_psi_BCS_n}
\psi^{\tilde{\lambda} = \tilde{l}, \, n \geq 1} (x_1, x_2) = & & \sum_{q=0}^{\floor{\frac{n+\tilde{l}-1}{2}}} c_{q}^{\tilde{\lambda} = \tilde{l}, \, n \geq 1} S_{q,n+\tilde{l}-q} (x_1, x_2) ,
\en
\end{small}

\noindent where $\tilde{l}=1,3,5,\ldots$, $\floor{x}$ denotes the floor function (i.e. the largest integer less than or equal to $x$), $S_{i,j} (x_1, x_2)$ are the Slater-like terms of Eq.~\eqref{eq_S}, and $c_{q}^{\tilde{\lambda} = \tilde{l}, \, n \geq 1}$ the coefficients given in Eq.~\eqref{eq_psi_BCS_n_app_coef}.  

The energy of $\psi^{\tilde{\lambda} = \tilde{l}, \, n} (x_1, x_2)$ is $ E/ \hbar \omega = \epsilon = n+\tilde{l}+1$. In this regime each energy leads to a subspace containing $\epsilon/2$ states for even $\epsilon$ and $(\epsilon-1)/2$ states for odd $\epsilon$. The number of Slater-like terms involved in the degenerated space with energy $\epsilon$ is $n/2$ for even $n$ and $(n+1)/2$ for odd $n$, being equivalent to the degeneracy of the energy $\epsilon$. The index of the involved Slater-like terms are all different implying that the expressions given by Eqs.~\eqref{eq_psi_BCS_n0} and \eqref{eq_psi_BCS_n} are the Slater-like decomposition of the state. Then, the number of available single-particle states ($g_{sp}$) in the degenerated space of energy $\epsilon$ is twice the number of involved Slater-like terms ($g_{sp}$ is $n$ for even $n$ and $n+1$ for odd $n$, or $\epsilon$ for even $\epsilon$ and $\epsilon-1$ for odd $\epsilon$). Therefore the entanglement of the states is bounded by the entanglement of the equally weighted sum of Slater-like states,  

\eq
\label{eq_SL_fermionization_bound}
\left(S_L^{\tilde{\lambda} = \tilde{l}, \, n}\right)^{bound} = 1- \frac{1}{\floor{\frac{\epsilon}{2}}} .
\en

\begin{figure}[tb]
\begin{center}
\includegraphics[height=0.35\columnwidth]{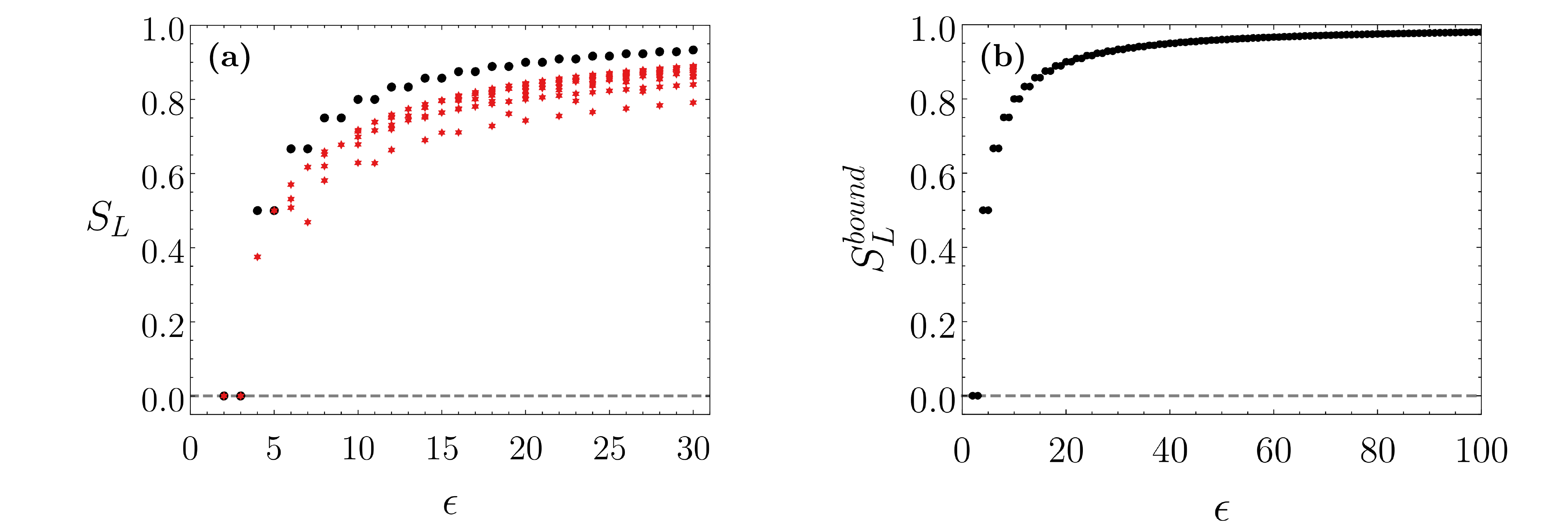}
\caption{Entanglement as a function of the total energy of the state for strongly repulsive interaction (red stars). The black dots give the upper bound of the entanglement, reached by the fifth excited state. Panel (b) shows this bound in a wider energy range. Notice that for a given energy $\epsilon \geq 2$, there are $\epsilon$ degenerated states for even $\epsilon$ and $\epsilon-1$ degenerated states for odd $\epsilon$.}
\label{fig_SL_fermionization}
\end{center}
\end{figure}

Figure \ref{fig_SL_fermionization} depicts the entanglement of the state $\psi^{\tilde{\lambda} = \tilde{l}, \, n} (x_1, x_2)$ calculated as $S_L = 1- 2\, \text{Tr} \rho^2$ in order to consider correlations beyond antisymmetry \cite{plastino_2009} (red stars), together with the previous entanglement bound (black dots) as a function of the energy of the states ($\epsilon \geq 2$). The first two states have zero entanglement, while the two states associated with $\epsilon = 5$ are the only ones that reach the upper bound. As expected, the entanglement is an increasing function of the energy. In contrast to the results obtained in the non-interacting regime (Sec. \ref{sec_unitarity}) these increasing correlations are provided by both, the center of mass and the relative wave-functions. We also notice that the states associated to the sub-space defined by even $\epsilon$ present a larger number of distinct entropy values, while the states with odd $\epsilon$ present more repeated (or less varied) entropy values. 

In the fermionization regime the atoms become impenetrable, their order along the longitudinal axis of the trap is fixed and therefore we can map directly the coordinates $x_{>}, x_{<}$ into the spatial ones. Rigourosly speaking, while both representations are equivalent due to the mapping between coordinates, in this regime the strict or rigid one-dimensional representation constitutes the more appropriate description for the system. In fact, this strict 1D behavior is the essential feature that allows to identify the spin orientation of the outermost atom in the trap in a tunneling experiment with few atoms \cite{murmann_2015}. In current experiments, the system is usually set up in the non-interacting ground state of the trap \cite{zurn_2012, serwane_2011}. Then, by increasing at a constant rate the magnetic field the system is ramped into the fermionization regime near the CIR. Our results show that entangled states could be experimentally obtained by reaching the strong repulsive regime in such a way. This results are quite different to the ones expected by considering the standard solutions that are in fact non-entangled states. For $\gamma \gg 1$, those standard solutions are obtained within the requirement of a finite energy which leads to solutions of Eq.~\eqref{eq_H_full} satisfying $\psi(r_1=r_2)=0$. In this case the last term of Eq.~\eqref{eq_H_full} vanishes, the variables decouple, and the mapping between variables $r_1$, $r_2$ and $r_>$, $r_<$ is direct. Therefore, for $\gamma \gg 1$, the states $S_{i,j} = [\phi_i ^{(1)} \phi_j^{2} -  \phi_j^{(1)} \phi_i^{(2)}]/\sqrt{2}$ (here we do not use coordinates in order to remark the mapping) are non-entangled eigen-functions with energy $E/\hbar\omega = \epsilon = i + j + 1$, where each energy leads to a subspace containing $\epsilon/2$ states $S_{i,j}$ for even $\epsilon$ and $(\epsilon-1)/2$ states $S_{i,j}$ for odd $\epsilon$. 

Summarizing, due to the mapping between coordinates all the results of the present section can be written in terms of $x_{1}, x_{2}$ instead of $x_{>}, x_{<}$, then the obtained states are proper Slaters following the Pauli exclusion principle.  A similar analysis to the one performed for the non-interacting regime holds. When preparing a state in the strong repulsive regime the obtained entanglement is null, while bringing a state from the interacting regime to the strongly repulsive regime by changing the scattering length near the CIR, leads to entangled states. The two degenerated excited states occupying the third energy level obtained with such a procedure would present the maximum entanglement. It is worth to notice that the absence of interaction is more selective than the strongly repulsive interaction in the sense that when reaching the non-interacting regime from the interacting regime there is only one state in which the system can land, while the strongly repulsive regime involves a degenerated sub-space in which any linear combination of the states is a possible state to land in. 


\section{Summary and Conclusions}
\label{sec_concl}

In the present work we revisited the entanglement and fermionization of two distinguishable harmonically confined fermions with zero-range interaction. A deeper understanding of entanglement and other correlations of the system helps to enhance its potentialities in quantum information processing. We present two alternative representations of the ground state which we associate with two different types of one-dimensional spaces. We found that the entanglement of the ground state is strongly conditioned by the one-dimensional space characteristics. In a strict one-dimensional space (one in which the particles can not pass through each other and the same particle is always confined to be for instance in the same left or right side) the particles have an intrinsic restriction which imposes an extra correlation leading to higher entanglement. On the other hand, in a one-dimensional space in which particles can occupy any arbitrary position this extra spurious correlation must be discarded leading to a smaller entanglement. Since the experimental realization of one-dimensional traps are actually three dimensional cigar-shaped potentials, we discussed a suitable definition of entanglement by properly analyzing which correlations must be considered. We also found that in the strongly attractive regime the state is a tightly localized state in the relative-coordinate leading to maximum entanglement. 

Our results indicate that entangled states could be obtained by experimentally reaching the non-interacting regime from the interacting one. Also, entangled states could be obtained when, by changing the scattering length near the CIR, a state in the interacting regime is brought into the strongly repulsive regime. These results are counterintuitive because in both cases the immediate or standard solutions of the non-interacting Hamiltonian or the strong repulsive one, consist in non-entangled states. In other words, the system selects entangled states instead of non-entangled states. Moreover, we showed that the first and third excited states obtained in the non-interacting and strongly repulsive regime respectively, are maximally entangled. We give the exact Schmidt decomposition of the ground and excited states in the absence of interactions, and the Slater decomposition of the ground and excited fermionized states in the strongly repulsive regime. From these decompositions we conclude that the absence of interaction is more selective than the strongly repulsive interaction in the sense that in the limit of vanishing interaction there is only one state in which the system can land, while the strongly repulsive regime involves a degenerated sub-space in which any linear combination of the states is also an eigenstate.

Our detailed analysis of the ground state shows that when reaching the non-interacting regime the available one-particle states turn off in a smooth way. The same happens for the available states in the strong repulsive regime characterized for the fermionization, the ground state changes smoothly from a superposition of Slater-like states to a finite superposition of Slaters. When building the many-particle state, this lack of accessible states lead to Pauli blocking as a strong signature of fermionization. Our findings set the needed stage for addressing the many-particle system within an alternative approach: the composite boson ansatz, which relies heavily upon a very good description and understanding of the two-particle physics. 

Finally, we would like to stress that all our analysis is valid for two either bosonic or fermionic distinguishable interacting particles. Due to its symmetry, the entire analysis of the ground state is also valid for two indistinguishable bosons. The experimental control of the interaction strength by means of Feshbach resonances requires particles of two different species (for example two different hyperfine states) meaning that when taking into account the experimental implementation the two considered particles must be distinguishable. Here, we focused on the fermionic case mainly motivated by the experiments of Refs. \cite{zurn_2012,rontani_2012}, in which the Pauli exclusion principle plays a crucial role in the preparation of the states \cite{serwane_2011}. It is important to keep in mind that due to the distinguishability of the particles, the symmetry of the wave-function arises as a consequence of the symmetry of the potential and not because of the symmetry imposed in the standard treatment of indistinguishability.



\ack{We are grateful to Andrea Vald\'es-Hern\'andez for fruitful discussions. We acknowledge grant BID-PICT 2017-2583 and GRFT-2018 MINCYT-C\'ordoba.}


\appendix

\section{Representations of the ground state} 
\label{sec_app_gs}

In the present Appendix we present details for the derivation of Eqs.~\eqref{eq_gs_P} and \eqref{eq_gs_S}. As mentioned in the main text we need to expand the parabolic cylinder function of Eq.~\eqref{eq_psi_r} as an Hermite series, then use some properties of the Hermite polynomials, and finally reconstruct the one-particle oscillator states $\phi_n(x)$.

There is a well-establish relationship between the parabolic cylinder functions $D_{\nu}(x)$ with non-negative integer index $\nu =n$ and the Hermite polynomials (see Ref. \cite{erdelyi_1985_book}),

\eq 
\label{eq_parabolic_to_hermite_app}
D_{n} (x) = e^{\frac{-x^2}{4}} \frac{ H_n \left( \frac{x}{\sqrt{2}} \right)}{2^{\frac{n}{2}}} .
\en

\noindent In addition, for positive real values of $x$, the parabolic cylinder function can be expressed as the following series \cite{erdelyi_1985_book}, 

\eq 
\label{eq_parabolic_series_P}
D_{\nu} (x) = \frac{2^{\frac{\nu}{2}}}{\Gamma\left(-\frac{\nu}{2}\right)} \sum_{n=0}^{\infty} \frac{(-1)^n D_{2n} (x)}{n!\, 2^n \left( n- \frac{\nu}{2} \right)} ,
\en

\noindent where $\Gamma$ denote the Gamma function as in the main text. It can also be expressed as

\eq 
\label{eq_parabolic_series_S}
D_{\nu} (x) = \frac{2^{\frac{\nu-1}{2}}}{\Gamma\left(\frac{1-\nu}{2}\right)} \sum_{n=0}^{\infty} \frac{(-1)^n D_{2n+1} (x)}{n!\, 2^n \left( n + \frac{1-\nu}{2} \right)} .
\en

\noindent These expansions of the parabolic cylinder can be done via explicit projection of $D_{\nu}(x)$ on $D_{n}(x)$, all the needed integrals are in Refs. \cite{abramowitz_stegun_1964_book, gradshteyn_2007_book}. Considering the relationship given by Eq.~\eqref{eq_parabolic_to_hermite_app} together with the symmetry of the Hermite polynomials it can be noticed that Eq.~\eqref{eq_parabolic_series_P} is a symmetric expansion in $x$, while Eq.~\eqref{eq_parabolic_series_S} is an antisymmetric expansion of $D_{\nu}$ in the same variable. 

Inserting Eq.~\eqref{eq_parabolic_to_hermite_app} in Eqs.~\eqref{eq_parabolic_series_P} and ~\eqref{eq_parabolic_series_S}, leads to expansions of the parabolic cylinder functions in terms of the Hermite polynomials. Since the variable of the wave-function of Eq.~\eqref{eq_psi_r} is positive ($\vert x_1 - x_2 \vert = x_> - x_<$ with $x_>$ ($x_<$) being Max$(x_1,x_2)$ (Min$(x_1,x_2)$)), the obtained expansions can be used. To separate the two variables involved in $\vert x_1 - x_2 \vert = x_> - x_<$, we make use of the addition formula

\eq 
\label{eq_hermite_x+y_app}
H_n \left( \frac{x + y}{\sqrt{2}} \right) = \sum_{k=0}^{n} \frac{1}{2^{\frac{n}{2}}} \binom{n}{k} H_k(x) H_{n-k} (y)  ,
\en

\noindent together with the explicit consideration of the symmetry properties of the Hermite polynomials, meaning that $H_{2n} (-x) = H_{2n} (x)$ and $H_{2n+1} (-x) = -H_{2n+1} (x)$, or more generally $H_{k} (-x) = (-1)^k H_{k} (x)$. 

When inserting Eq.~\eqref{eq_parabolic_to_hermite_app} in Eq.~\eqref{eq_parabolic_series_P} all the involved Hermite polynomials are a sum of even powers of the variable. Taking into account that $\vert x_1 - x_2 \vert^{2n} = (x_1 - x_2)^{2n}$, using Eq.~\eqref{eq_hermite_x+y_app}, writing the Hermite polynomials in terms of the one-particle oscillator states $\phi_n(x)$, and regrouping terms one is able to obtain the representation of the ground state as a sum of Permanents (P) and product terms thus leading to a bosonic-like representation of the ground state

\begin{eqnarray}
\label{eq_gs_P_app}
\psi_{gs} (x_1, x_2) =  & & \sum_{n=1}^{\infty} \sum_{k=0}^{n-1} c^P(n,k,\tilde{\lambda}) P_{k, 2n-k} (x_1, x_2) 
\nonumber \\
& & +\sum_{n=0}^{\infty} c(n,\tilde{\lambda}) \phi_n(x_1) \phi_n(x_2) ,
\end{eqnarray}

\noindent with the coefficients given by

\eq
\label{eq_c_P_app}
c^P(n,k,\tilde{\lambda}) = \sqrt{2}\, c(n,\tilde{\lambda})\, \frac{n! (-1)^{k+n}}{\sqrt{(2n-k)! k!}},
\en

\noindent where

\begin{eqnarray}
\label{eq_c_app}
c(n,\tilde{\lambda}) = & &  \frac{\tilde{\lambda} 2^{\frac{\tilde{\lambda}+1}{2}}}{\Gamma(1-\frac{\tilde{\lambda}}{2})} \sqrt{\frac{\Gamma(-\tilde{\lambda})}{\Psi \left( \frac{1-\tilde{\lambda}}{2} \right) -\Psi \left( - \frac{\tilde{\lambda}}{2} \right) }} 
\, \frac{1}{\tilde{\lambda}-2n} \frac{(2n-1)!!}{2^n n!} ,
\end{eqnarray}

\noindent and the Permanents are indicated as

\eq
\label{eq_P_app}
P_{k, 2n-k} (x_1, x_2) = \frac{\phi_k(x_1) \phi_{2n-k}(x_2) + \phi_{2n-k}(x_1) \phi_k(x_2) }{\sqrt{2}} .
\en

Inserting Eq.~\eqref{eq_parabolic_to_hermite_app} in Eq.~\eqref{eq_parabolic_series_S}, and repeating the previous procedure it is possible to obtain the fermionic-like representation of the ground state as a sum of Slater-like (S) terms

\eq
\label{eq_gs_S_app}
\psi_{gs} (x_1, x_2) =  & & \sum_{n=0}^{\infty} \sum_{k=0}^{n} c^S(n,k,\tilde{\lambda}) S_{k, 2n+1-k} (x_1, x_2) ,
\en

\noindent where the coefficient are   

\begin{eqnarray}
\label{eq_z_S_app}
c^S(n,k,\tilde{\lambda}) =& & \frac{ 2^{\frac{\tilde{\lambda}}{2}+1}}{\Gamma(\frac{1-\tilde{\lambda}}{2})} \sqrt{\frac{\Gamma(-\tilde{\lambda})}{\Psi \left( \frac{1-\tilde{\lambda}}{2} \right) -\Psi \left( - \frac{\tilde{\lambda}}{2} \right) }} 
\nonumber \\
& &\times  \frac{1}{\tilde{\lambda}-(2n+1)} \frac{(2n+1)!!}{2^n} \frac{(-1)^{n+1+k}}{\sqrt{(2n+1-k)! k!}} ,
\end{eqnarray}

\noindent and the Slater-like terms are

\eq
\label{eq_S_app}
S_{k, 2n+1-k} (x_1, x_2) = \frac{\phi_k(x_<) \phi_{2n+1-k}(x_>) - \phi_{2n+1-k}(x_<) \phi_k(x_>) }{\sqrt{2}}.
\en

As it is mentioned in the main text, the ground state is obtained by setting $n=0$ in Eq.~\eqref{eq_psi_R} and $-\infty < \tilde{\lambda} \leq 1$ in Eq.~\eqref{eq_psi_r}. The obtained state is a symmetric state under particle exchange, therefore, both representations have this symmetry. The Slater-like terms in Eq.~\eqref{eq_S_app} are not antisymmetric but symmetric under particle exchange, reason why we refer to these states as Slater-like terms and not as proper Slaters.

For the sake of completeness we would like to mention that in case of being interested in obtaining expansions of the excited states with $n \neq 0$ in Eq.~\eqref{eq_psi_R} in terms of the one-particle oscillator functions, one should also use Eq.~\eqref{eq_hermite_x+y_app} in Eq.~\eqref{eq_psi_R}. Then one should multiply that expression with the one obtained when inserting Eq.~\eqref{eq_parabolic_to_hermite_app} in Eq.~\eqref{eq_parabolic_series_S} or Eq.~\eqref{eq_parabolic_series_S} followed by the use of Eq.~\eqref{eq_hermite_x+y_app}. After that one should group the Hermite polynomials depending on the same variables and use

\eq 
\label{eq_hermite_contraction_app}
H_n (x) H_m(x) = \sum_{k=0}^{\text{Min}(n,m)} 2^k k! \binom{m}{k} \binom{n}{k} H_{m+n-2k}(x)  ,
\en

\noindent in order to obtain a single Hermite polynomial in each variable and finally be able to reconstruct the one-particle oscillator functions.

\section{States in the Deep Repulsive Regime} 
\label{sec_app_slaters}

In the present section we derive the expressions used in Sec. \ref{sec_fermionization}. Let us start by considering a general state with $\tilde{\lambda} = \tilde{l}$ with integer $\tilde{l}=0,1,2,\ldots$. Then, the total wave-function can be written as

\begin{small}
\begin{eqnarray}
\label{eq_psi_gen_app}
\psi^{\tilde{\lambda} = \tilde{l}, \, n} (x_1, x_2) = & & \sqrt{\frac{m \omega }{\hbar \pi}} \frac{e^{-\frac{x_1^2+x_2^2}{2}}}{\sqrt{2^{(\tilde{l}+n)} \tilde{l}! n!}} 
\, H_n \left( \frac{x_1+x_2}{\sqrt{2}} \right) H_{\tilde{l}} \left( \frac{\vert x_1-x_2 \vert}{\sqrt{2}} \right) .
\end{eqnarray}
\end{small}

\noindent Expanding the Hermite polynomials and reconstructing the one-particle oscillator states $\phi_n(x)$, these states give

\begin{small}
\begin{eqnarray}
\label{eq_psi_gen_app_prod}
\psi^{\tilde{\lambda} = \tilde{l}, \, n} (x_1, x_2) = & & \sqrt{\frac{\tilde{l}! n!}{2^{(\tilde{l}+n} }} \, \sum_{k=0}^{n} \, \sum_{l=0}^{\tilde{l}} \, \sum_{r=0}^{\text{Min}(k,l)} \, \sum_{p=0}^{\text{Min}(n-k,\tilde{l}-l)} (-1)^{\tilde{l}-l}
\nonumber \\
& & \times c^{\tilde{l}, \, n}(k,l,r,p) \, \phi_{k+l-2r} (x^>) \, \phi_{n+\tilde{l}-k-l-2p} (x^<) ,
\end{eqnarray}
\end{small}

\noindent with 

\begin{small}
\eq
\label{eq_psi_gen_app_prod_coef}
c^{\tilde{l}, \, n}(k,l,r,p) = \frac{\sqrt{(k+l-2r)!(n-k+\tilde{l}-l-2p)!}}{(k-r)!(l-r)!(n-k-p)!(\tilde{l}-l-p)!r!p!} .
\en
\end{small}

\noindent Taking into account that the values of $\tilde{\lambda}$ of our interest are odd, the above expression is equivalent to

\begin{small}
\eq
\label{eq_psi_BCS_n0_app}
\psi^{\tilde{\lambda} = \tilde{l}, \, n=0} (x_1, x_2) = & & \sqrt{\frac{\tilde{l}!}{2^{(\tilde{l}-1)}}} \, \sum_{l=0}^{\frac{\tilde{l}-1}{2}} \frac{(-1)^l}{\sqrt{(\tilde{l}-l)!l!}} S_{l,\tilde{l}-l} (x_1, x_2) ,
\en
\end{small}

\noindent and

\begin{small}
\eq
\label{eq_psi_BCS_n_app}
\psi^{\tilde{\lambda} = \tilde{l}, \, n \geq 1} (x_1, x_2) = & & \sum_{q=0}^{\floor{\frac{n+\tilde{l}-1}{2}}} c_{q}^{\tilde{\lambda} = \tilde{l}, \, n \geq 1} S_{q,n+\tilde{l}-q} (x_1, x_2) ,
\en
\end{small}

\noindent where $\floor{x}$ denotes the floor function i.e. the largest integer less than or equal to $x$, $S_{i,j} (x_1, x_2)$ are the Slater-like terms of Eq.~\eqref{eq_S}, and

\begin{small}
\begin{eqnarray}
\label{eq_psi_BCS_n_app_coef}
\hspace{-0.5cm} c_{q}& &^{\tilde{\lambda} = \tilde{l}, \, n \geq 1} =  
\sqrt{\frac{\tilde{l}! \,n!}{2^{(\tilde{l}+n-1)} }} \sum_{l=0}^{\frac{\tilde{l}-1}{2}} (-1)^l 
\left\lbrace \sum _{k=0}^{\bar{n}} \left(  
\sum _{i = \left| k-l\right| }^{k+l} \,
\sum _{j = \left| (n-k)-(\tilde{l} -l)\right| }^{(n-k)+(\tilde{l} -l)} \right. \right.
\nonumber\\
& & \,\, c^{\tilde{l}, \, n} \left( k,l,\frac{(k+l-i)}{2},\frac{(n-k)+(\tilde{l}-l)-j}{2} \right)
 (\delta_{i,q} \delta_{j,n+\tilde{l}-q}- \delta_{j,q} \delta_{i,n+\tilde{l}-q}) 
\nonumber\\
& & + \sum _{i = \left|(n-k)-l \right| }^{(n-k)+l} \,
\sum _{j = \left| k-(\tilde{l} -l)\right| }^{k+(\tilde{l} -l)} 
 c^{\tilde{l}, \, n} \left( k, \tilde{l}-l, \frac{k+(\tilde{l}-l)-j}{2},\frac{(n-k)+l-i}{2} \right)
\nonumber\\ 
& & \times \left(\delta_{i,q} \delta_{j,n+\tilde{l}-q} - \delta_{j,q} \delta_{i,n+\tilde{l}-q} \right) \left. \vphantom{\left(  
\sum_{i = \left| k-l\right| }^{k+l} \sum_{j = \left| (n-k)-(\tilde{l} -l)\right| }^{(n-k)+(\tilde{l} -l)} \right.} \right) + 
 p(n) \sum _{i= \left| \frac{n}{2}-l \right| }^{\frac{n}{2} + l} \,
 \sum _{j= \left| \frac{n}{2}- (\tilde{l} -l) \right| }^{\frac{n}{2} + (\tilde{l} -l) }
\nonumber\\
 & & \,\, c^{\tilde{l}, \, n} \left( \frac{n}{2},\tilde{l}-l,\frac{\frac{n}{2} + (\tilde{l} -l) -j}{2} ,\frac{\frac{n}{2}+l-i}{2} \right)
 \left. \left( \delta_{i,q} \delta_{j,n+\tilde{l}-q } - \delta_{j,q} \delta_{i,n+\tilde{l}-q} \right) 
\vphantom{\left\lbrace \sum _{k=0}^{\bar{n}} \left(  
\sum_{i = \left| k-l\right| }^{k+l} 
\sum_{j = \left| (n-k)-(\tilde{l} -l)\right| }^{(n-k)+(\tilde{l} -l)} \right. \right.} \!\right\rbrace ,
\end{eqnarray}
\end{small}

\noindent with $\bar{n} = n/2 -1$ and $p(n)=1$ for even $n$, while $\bar{n} = (n-1)/2$ and $p(n)=0$ for odd $n$. 



\section*{References}

\bibliography{bib_fermionization}

\bibliographystyle{iopart-num}


\end{document}